\documentclass{article}

\usepackage{spconf,amsmath,graphicx} 
\usepackage[hidelinks]{hyperref}

\usepackage[caption=false,font=footnotesize]{subfig}

\usepackage[acronym]{glossaries} 
\usepackage{xcolor} 
\usepackage{amsmath, amssymb} 
\usepackage{multirow} 
\usepackage{comment} 
\usepackage[tracking=true]{microtype} 
\usepackage{graphicx} 
\usepackage{balance} 

\usepackage{booktabs}   
\usepackage{caption}    

\newacronym{3gpp}{3GPP}{3rd Generation Partnership Project}
\newacronym{5gnr}{5G NR}{5G New Radio}
\newacronym{ap}{AP}{antenna point}
\newacronym{bie}{BIE}{best integer equivariant}
\newacronym{cf}{CF}{cell-free}
\newacronym{cnn}{CNN}{convolutional neural network}
\newacronym{conv2d}{Conv2D}{2D-Convolutional}
\newacronym{cpp}{CPP}{carrier phase positioning}
\newacronym{csi}{CSI}{channel state information}
\newacronym{dmimo}{D-MIMO}{distributed multiple-input multiple-output}
\newacronym{dl}{DL}{deep learning}
\newacronym{dlmf}{DLMF}{distributed location management function}
\newacronym{dmrs}{DMRS}{demodulation reference signal}
\newacronym{dnn}{DNN}{deep neural network}
\newacronym{ecdf}{ECDF}{empirical cumulative distribution function}
\newacronym{flop}{FLOP}{floating-point operation}
\newacronym{fc}{FC}{fully connected}
\newacronym{gd}{GD}{gradient descent}
\newacronym{gnss}{GNSS}{global navigation satellite system}
\newacronym{lmf}{LMF}{location management function}
\newacronym{los}{LoS}{line-of-sight}
\newacronym{mimo}{MIMO}{multiple-input multiple-output}
\newacronym{ml}{ML}{machine learning}
\newacronym{mle}{MLE}{maximum likelihood estimation}
\newacronym{mlp}{MLP}{multi-layer perceptron}
\newacronym{mse}{MSE}{mean-squared error}
\newacronym{nn}{NN}{neural network}
\newacronym{prs}{PRS}{positioning reference signal}
\newacronym{psd}{PSD}{power spectral density}
\newacronym{relu}{ReLU}{Rectified Linear Unit}
\newacronym{rmse}{RMSE}{Root mean-squared error}
\newacronym{snr}{SNR}{signal-to-noise ratio}
\newacronym{srs}{SRS}{sounding reference signal}
\newacronym{tdoa}{TDoA}{time-difference-of-arrival}
\newacronym{toa}{ToA}{time-of-arrival}
\newacronym{tof}{ToF}{time-of-flight}
\newacronym{ue}{UE}{user equipment}

\title{
Phase-Only Positioning in Distributed MIMO Under Phase Impairments: \\ AP Selection using Deep Learning}
\name{%
\begin{tabular}{@{}c@{}}
Fatih Ayten$^{1}$, Musa Furkan Keskin$^{2}$, Akshay Jain$^{3}$, Mehmet C. Ilter$^{1}$, Ossi Kaltiokallio$^{1}$,\\
Jukka Talvitie$^{1}$, Elena Simona Lohan$^{1}$, Mikko Valkama$^{1}$
\end{tabular}%
}

\address{%
$^{1}$Department of Electrical Engineering, Tampere Wireless Research Center, \\ Tampere University, Finland\\
$^{2}$Department of Electrical Engineering, Chalmers University of Technology, Sweden\\
$^{3}$Radio Systems Research, Nokia Bell Labs, Espoo, Finland%
}
\begin{document}
%
\maketitle
\begin{abstract}
\Gls{cpp} can enable cm-level accuracy in next-generation wireless systems, while the recent literature shows that accuracy remains high using \emph{phase-only measurements} in the emerging \gls{dmimo} networks. However, the impact of phase synchronization errors on such systems remains insufficiently explored. To address this gap, we first show that the proposed hyperbola intersection method achieves highly accurate positioning even in the presence of phase synchronization errors when trained on appropriate data reflecting such impairments. We then introduce a \gls{dl}–based \gls{dmimo} \gls{ap} selection framework that ensures high-precision localization under phase synchronization errors. Simulation results show that the proposed framework improves positioning accuracy compared to prior art methods, while reducing inference complexity by approximately $19.7\%$.


\end{abstract}
\glsresetall   
\vspace{-1.5em}
\section{Introduction}\vspace{-3mm}
Carrier phase–only positioning represents a distinct approach where localization is performed exclusively using carrier phase observations, without relying on complementary measurements such as \gls{toa} or \gls{tdoa}. Unlike conventional hybrid schemes that treat carrier phase as an auxiliary source of information, this paradigm leverages the fine-grained phase data as the sole input for estimating the \gls{ue} position \cite{Cha2025,10475845,5g_carrier_phase_prs_1}. Such a setting introduces both opportunities for achieving sub-centimeter accuracy and challenges due to the inherent integer ambiguity problem \cite{10536135,10232971} and sensitivity to hardware impairments \cite{11100563}. Recent studies in distributed and phase-coherent \gls{mimo} systems have begun exploring this concept, highlighting its potential for next-generation cellular positioning \cite{10536135,cell_free_massive_mimo}.

However, practical \gls{dmimo} deployments are inherently affected by phase synchronization errors across distributed \glspl{ap}, which significantly degrade performance \cite{larsson_mimo}. In addition, a wide range of applications in next-generation wireless networks must satisfy strict latency requirements \cite{10597064}, which can be alleviated by reducing model complexity. 

Motivated by the aforementioned challenges, in this paper, we propose a learning-based \gls{lmf} that leverages \glspl{dnn} to maintain robustness against such impairments. Furthermore, our framework incorporates adaptive \gls{ap} selection strategies, leveraging received signals and deployment geometry. Extensive numerical evaluations confirm that the proposed \gls{dl}-based \gls{dlmf} achieves high-precision localization in realistic 6G deployment scenarios, even under challenging phase synchronization errors, with reduced complexity.

\begin{figure}[t!] 
    \centering
    \includegraphics[width=0.48\textwidth]{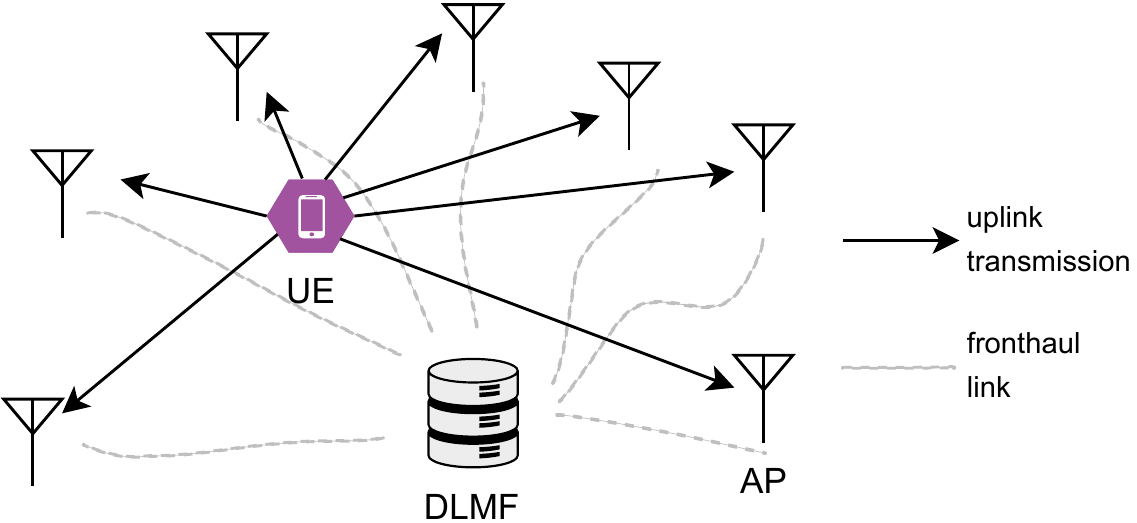}
    \caption{Uplink UE positioning in a distributed AP deployment, where only carrier phase measurements are taken into account for positioning.}\vspace{-3mm}
    \label{fig:system_model}
\end{figure}

\vspace{-3mm}\section{System Architecture} \vspace{-3mm}

The \gls{lmf} is the key enabler for positioning purposes in \gls{5gnr}, and it can be used to determine the position of a \gls{ue} utilizing measurements \cite{lmf}. The \gls{dlmf} extends the traditional LMF by moving certain positioning tasks closer to the \gls{ue}, reducing latency and improving accuracy \cite{US20200367022A1}. While a central LMF may still coordinate overall operations, the DLMF boosts performance through localized processing. In this work, we assume a \gls{dmimo} system for \gls{ue} positioning, coordinated by a \gls{dlmf} unit, as shown conceptually with seven \glspl{ap} in Fig.~\ref{fig:system_model}.

Similar to \cite{pimrc_main_paper,pimrc_workshop_paper,10283650}, we study an uplink configuration involving $I$ spatially distributed \glspl{ap}. 
The \gls{ue} is located at an unknown position $\boldsymbol{x}_{\text{ue}} \in \mathbb{R}^2$, while the positions of the \glspl{ap} are known and denoted as $\boldsymbol{x}_{\text{ap,}i} \in \mathbb{R}^2$ for $i \in \{0,\dots,I-~1\}$. Our analysis focuses on \gls{los} propagation. Extensions to multipath environments and three-dimensional positioning remain important directions for future work. 

The \gls{ue} transmits a narrowband pilot signal. 
The phase of the received signal at the $i$-th \gls{ap} is modeled as $\theta_i = -\frac{2\pi}{\lambda} d_i + \gamma_i + \phi_{\text{ue}} + 2 \pi z_i + n_i,$ where $\lambda$ is the wavelength, 
$d_i = \Vert \boldsymbol{x}_{\text{ue}} - \boldsymbol{x}_{\text{ap,}i} \Vert$ is the Euclidean distance, 
$\gamma_i \sim \mathcal{N}(0,\sigma^2)$ represents the phase perturbation between the \gls{dlmf} and the $i$-th \gls{ap}, 
$\phi_{\text{ue}}$ is the phase offset between the \gls{ue} and the \gls{dlmf}, 
$z_i \in \mathbb{Z}$ is the integer ambiguity, 
and $n_i \sim \mathcal{N}(0,\nu_i^2)$ models the phase measurement noise. We assume $\gamma_i$ and $n_i$ are independent.

One \gls{ap} (e.g., $i=0$) is designated as the reference by the network. 
The corresponding differential measurements for $m \in \{1,\dots,I-1\}$ are then defined as\vspace{-2mm}
\begin{equation}\label{eq:diff_meas}
    \delta_m = -\frac{\lambda}{2\pi}\bigl(\theta_m - \theta_0\bigr) = \Delta d_m + \Delta \gamma_m - \lambda \Delta z_m + \Delta n_m
\end{equation}
where $\Delta d_m = d_m - d_0$, 
$\Delta \gamma_m = (\lambda/2\pi)(\gamma_0 - \gamma_m)$, 
$\Delta z_m =~ z_m - z_0$, 
and $\Delta n_m = (\lambda/2\pi)(n_0 - n_m)$. 
Stacking the differential measurements in \eqref{eq:diff_meas} yields 
$\boldsymbol{\delta} = [\delta_1,\dots,\delta_{I-1}]^\top$. 
Each $\delta_m$ geometrically corresponds to a set of hyperbolas parameterized by $\Delta z_m$, 
with foci located at $\boldsymbol{x}_{\text{ap,}m}$ and $\boldsymbol{x}_{\text{ap,}0}$.

\begin{figure}[t!]
    \centering
    \vspace{2mm}
    \includegraphics[width=0.48\textwidth]{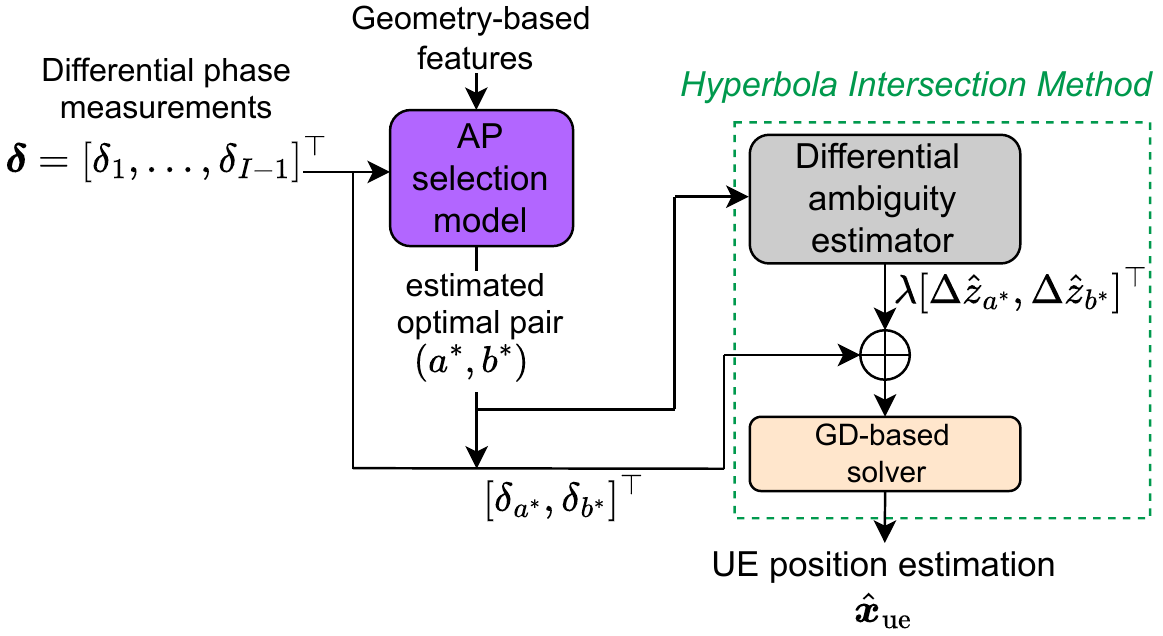}
    \vspace{-7mm}
    \caption{Block diagram of the proposed phase-only positioning approach.}\vspace{-5mm}
    \label{fig:proposed_approach} 
\end{figure}

\vspace{-2mm}
\section{Phase-only Positioning}\vspace{-0.5em}
In this section, we introduce phase-only positioning with AP selection, which integrates \gls{dl} and \gls{gd} techniques to perform positioning. The overall framework is illustrated in Fig.~\ref{fig:proposed_approach}. The hyperbola intersection method, originally introduced in \cite{pimrc_workshop_paper}, is adopted as the underlying localization model. The proposed \gls{dl}-based \gls{ap} selection approach identifies the optimal pair of ambiguities that minimizes the \gls{ue} positioning error.

\vspace{-3mm}\subsection{Hyperbola Intersection Method}\vspace{-1mm}
\textls[-12]{This method comprises two primary components: (1) a \textit{differential ambiguity estimator} implemented using a \gls{mlp}, and (2) a \textit{\gls{gd}-based solver}. Due to space limitations, a detailed description of the method is omitted; readers are referred to Sec.~III of~\cite{pimrc_workshop_paper} for a similar underlying approach.} 

The \textit{differential ambiguity estimator} produces the vector of estimates  
$\Delta\hat{\boldsymbol{z}} = \left[\Delta\hat{z}_{j_1}, \dots, \Delta\hat{z}_{j_{\lvert\mathcal{J}\rvert}} \right]^\top$  
from the differential measurements $\boldsymbol{\delta}$, where  
$\mathcal{J} = \{ j_k \}_{k=1}^{\lvert\mathcal{J}\rvert} \subseteq \{1,2,\dots,I-1\}$  
denotes the set of AP indices for which differential ambiguities are estimated, and  
$\lvert \cdot \rvert$ is the cardinality.  
Here, $\Delta\hat{z}_{j_k}$ represents the estimate of $\Delta z_{j_k}$. The total number of possible labels across all differential ambiguities is expressed as $Q$, which depends on the \gls{ap} positions. 
The estimated ambiguities $\Delta\hat{\boldsymbol{z}}$, scaled by $\lambda$, are added to the differential phase measurements $\boldsymbol{\delta}$ to obtain the differential distance estimates $\Delta\hat{\boldsymbol{d}}$.

The \textit{\gls{gd}-based solver} iteratively refines the \gls{ue} position estimate using differential distance estimates $\Delta\hat{\boldsymbol{d}}$. At each step, differential distances at the current estimate are computed. A quadratic cost function is then evaluated, and the position is updated with a learning rate of~$\alpha$. After $T$ iterations, the final position estimate $\hat{\boldsymbol{x}}_{\text{ue}}$ is obtained.

\vspace{-3mm}\subsection{Proposed AP Selection Method}\vspace{-1mm}
For any unordered pair of differential ambiguities $({\Delta z_a, \Delta z_b})$ with $a,b \in \{1, \ldots, I-1\}$ and $a \neq b$, the intersection of the corresponding hyperbolas yields a unique \gls{ue} position estimate. As shown in \cite{pimrc_workshop_paper}, the computational complexity of the hyperbola intersection method grows with the number of estimated ambiguities, making two-ambiguity localization, i.e., $\lvert \mathcal{J} \rvert=2$, computationally more efficient. The operation of estimating only two of all possible ambiguities is performed as follows: the shared layers of the ambiguity estimator model in \cite{pimrc_workshop_paper} remain active, and only the parallel branches corresponding to those ambiguities are activated, since each branch of the model produces an ambiguity estimate. However, the optimal pair $({\Delta z_a, \Delta z_b})$ may depend on factors such as received \gls{snr} and AP geometry. To address this, we propose a \gls{nn}-based approach to identify the pair that maximizes positioning accuracy.

We consider two classes of features: measurement-based and geometry-based. \textit{Measurement-based features} include (1) the differential phase measurements $\boldsymbol{\delta}$ and (2) the \gls{snr} values (in dB) at each \gls{ap}. Here, \gls{snr} quantifies only measurement noise and excludes phase perturbation variance. To ensure consistent feature dimensionality, an additional zero-valued element is appended to $\boldsymbol{\delta}$ for the reference \gls{ap}, since $\boldsymbol{\delta}$ is defined only for $m \in \{1,\dots,I-1\}$. \textit{Geometry-based features}, in turn, capture deployment-specific characteristics of the \gls{ap} configuration. For each \gls{ap} in the system, we extract the following geometric relationships: (1) distance to the reference \gls{ap}, (2) the angle between the current \gls{ap} and the reference \gls{ap}, (3) index of the closest neighbor \gls{ap}, (4) the angle between the current \gls{ap} and its closest neighbor \gls{ap}, (5) index of the farthest \gls{ap}, and (6) the angle between the current \gls{ap} and the farthest \gls{ap}. For each \gls{ap}, the measurement-based and geometry-based features are concatenated into a vector of dimension $8$. The per-\gls{ap} feature vectors are then concatenated to form the overall feature vector $\boldsymbol{F} \in \mathbb{R}^{8I}$. 

\vspace{-0mm}We employ a \gls{mlp}-based \gls{nn} to predict the positioning error associated with every ambiguity pair. Let $e_{a,b}$ denote the ground-truth positioning error computed for the ambiguity pair $(\Delta z_a,\Delta z_b)$ using a pre-trained hyperbola intersection model.  
The network takes the feature vector $\boldsymbol{F}$ as input and predicts these errors via $\hat{\boldsymbol{e}} \;=\; f_{\boldsymbol{\theta}}(\boldsymbol{F}) \;\in\; \mathbb{R}^\eta,$ where $f_{\boldsymbol{\theta}}(\cdot)$ denotes the mapping implemented by the \gls{nn} with trainable parameters $\boldsymbol{\theta}$, and each element $\hat{e}_{a,b}$ corresponds to the predicted positioning error for $(\Delta z_a,\Delta z_b)$. Here, $\eta = \binom{I-1}{2}$ denotes the total number of pairs that can be formed from $I-1$ ambiguities. The optimal pair for a given input is then selected as $(a^\star,b^\star) \;=\; \arg\min_{(a,b)} \; \hat{e}_{(a,b)}$. The \gls{nn} architecture consists of an input layer of size $8I$, followed by five hidden layers with sizes $A$, $2A$, $4A$, $2A$, and $A$, respectively, and concludes with an output layer of dimension $\binom{I-1}{2}$. The input and hidden layers employ the \gls{relu} activation function, while the output layer uses a linear activation. Given a training set $\{(\boldsymbol{F}^{(i)}, \boldsymbol{e}^{(i)})\}_{i=1}^N$ with $N$ samples, where $\boldsymbol{e}^{(i)}$ contains the ground-truth positioning errors for all pairs, the network is trained by minimizing the \gls{mse} loss function expressed as $\frac{1}{N}\sum_{i=1}^{N} \Vert \hat{\boldsymbol{e}}^{(i)} - \boldsymbol{e}^{(i)} \Vert^2.$

\vspace{-3mm}\section{Inference Complexity}\vspace{-2mm}
We next evaluate the inference complexities of the proposed \gls{ap} selection method and the hyperbola intersection approach. For a fair comparison with~\cite{pimrc_workshop_paper}, the same complexity metric is adopted—namely, the \gls{flop} count—where each elementary arithmetic operation is counted as one \gls{flop}.

A \gls{fc} layer with input and output sizes $n_i$ and $n_o$ requires approximately $2n_o n_i$ \glspl{flop}. Since the proposed \gls{ap} selection model consists solely of \gls{fc} layers, its complexity is $\mathcal{C}_{\text{APS}} \approx 2A(20A+8I+\eta)$~\glspl{flop}. As reported in~\cite{pimrc_workshop_paper}, the hyperbola intersection method has complexity $\mathcal{C}_{\text{HI}} = \mathcal{C}_{\text{DAE}} + \mathcal{C}_{\text{GD}}$, where the differential ambiguity estimator incurs $\mathcal{C}_{\text{DAE}} \approx D^2(4|\mathcal{J}|+16)+D(2Q+4I)$ \glspl{flop}, and the \gls{gd}-based solver requires $\mathcal{C}_{\text{GD}} \approx T(18|\mathcal{J}|+10)$ \glspl{flop}. When estimating only two ambiguities rather than all, both $|\mathcal{J}|$ and $Q$ decrease, which reduces $\mathcal{C}_{\text{DAE}}$ and $\mathcal{C}_{\text{GD}}$. 
However, for the overall complexity to be reduced, this decrease must outweigh the additional cost of $\mathcal{C}_{\text{APS}}$.

\vspace{-2mm}
\section{Simulation Results}\vspace{-2mm}
In our simulations, a $100\,\text{m}^2$ square region is used as the evaluation area, comprising $I=9$ distributed \glspl{ap} arranged to emulate a \gls{dmimo} network topology. The \gls{ap} locations are identical to those in~\cite{pimrc_workshop_paper}, resulting in a total of $Q = 334$ possible output classes for all differential ambiguities. When the \gls{ap} selection model is active and only two ambiguities are estimated, $Q$ varies with the chosen ambiguity pair. To account for this, $Q/4$ is used, with the factor of $4$ representing $\frac{I-1}{2}$. The \gls{5gnr} \gls{srs}-based uplink pilot configuration and noise settings are identical to those in~\cite{pimrc_workshop_paper}, while only a $0\,\text{dBm}$ uplink transmit power level is considered in this work. 
The \gls{gd}-based solver is set to perform $T=500$ iterations using a learning rate of $\alpha = 0.08$.

\vspace{-3mm}\subsection{Training of Proposed Neural Networks}\vspace{-2mm}
\textls[-3]{The supervised training procedure involves two steps. First, we train the differential ambiguity estimator in the hyperbola intersection method to investigate the effect of the phase perturbations on the ambiguity estimation accuracy and to produce the correct ambiguity pair $(\Delta z_a,\Delta z_b)$ for each training sample of the \gls{ap} selection network. During the training of the differential ambiguity estimator, all parallel branches are activated, i.e., no \gls{ap} selection is applied. The neuron parameter is set to $D = 150$. This value is increased from $128$ in~\cite{pimrc_workshop_paper} to $150$ to preserve high ambiguity estimation accuracy in the presence of phase disturbances. The estimator is trained using $700 \times 10^3$ samples for training and $150 \times 10^3$ samples for validation, with \gls{ue} positions, noise, and phase perturbations randomly sampled for each instance. 

Second, we train the \gls{ap} selection network. The neuron parameter is set to $A = 100$. Training of the \gls{ap} selection model requires computing the positioning error for every ambiguity pair $(\Delta z_a,\Delta z_b)$ using the pre-trained hyperbola intersection model. Specifically, for each training instance, a random \gls{ue} location, noise and phase perturbation realizations are generated first. The corresponding feature matrix $\boldsymbol{F}$ is then constructed. For each ambiguity pair, the hyperbola intersection method is applied to obtain the \gls{ue} positioning error, which serves as the ground-truth label for supervised learning. Using this procedure, $200 \times 10^3$ samples are employed for training, $40 \times 10^3$ for validation, and another $40 \times 10^3$ for testing.}

\vspace{-2mm}\subsection{Inference Complexity Reduction}\vspace{-2mm}
\textls[-4]{With the considered parameters, the inference complexity of the \gls{ap} selection model is
$\mathcal{C}_{\text{APS}} \approx 0.420 \times 10^6$~\glspl{flop}.
For the hyperbola-intersection model, the inference complexity is
$\mathcal{C}_{\text{HI,all}} \approx 1.262 \times 10^6$~\glspl{flop} when all ambiguities are used, and
$\mathcal{C}_{\text{HI,2}} \approx 0.594 \times 10^6$~\glspl{flop} when only two ambiguities are used. By using two ambiguities in combination with the \gls{ap} selection model instead of all ambiguities, 
the overall positioning complexity is reduced by approximately $19.7\%$}.

\begin{figure*}[t!]
    \centering

    \subfloat[\label{fig:acc_and_95th_error}]{
        \includegraphics[width=0.32\textwidth]{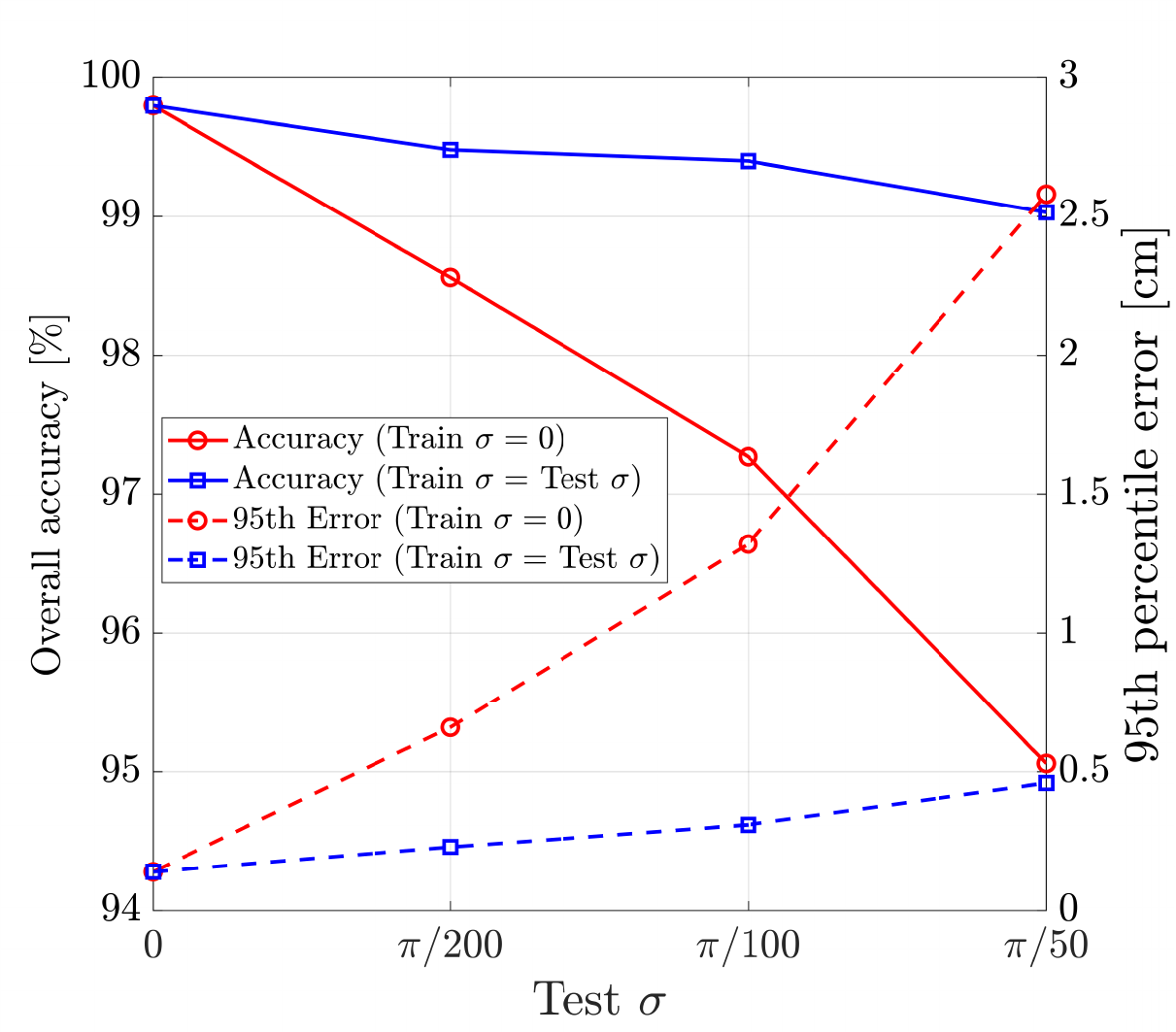}
    }\hfill
    \subfloat[\label{fig:positioning_ecdf}]{
        \includegraphics[width=0.32\textwidth]{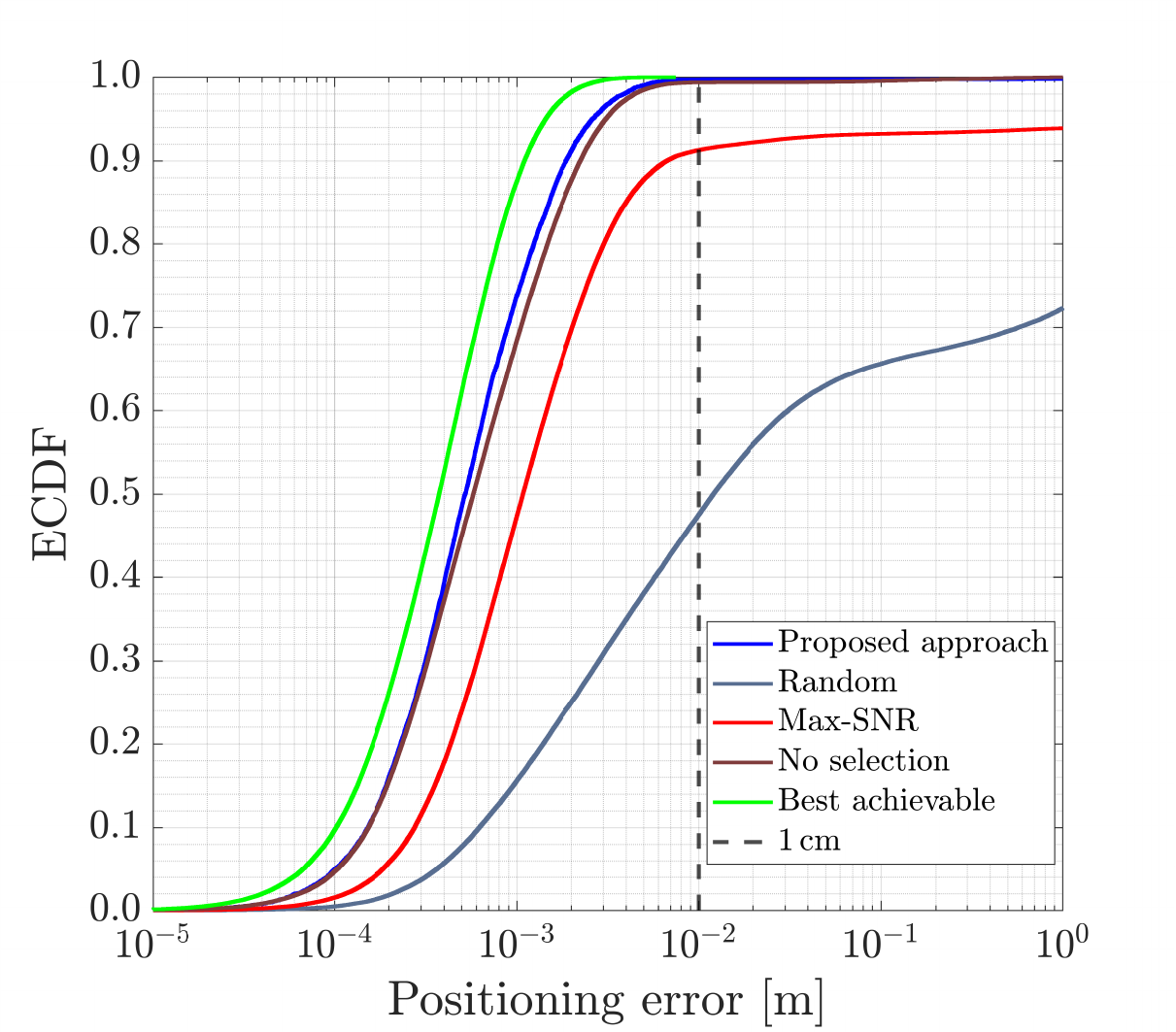}
    }\hfill
    \subfloat[\label{fig:positioning_ccdf}]{
        \includegraphics[width=0.32\textwidth]{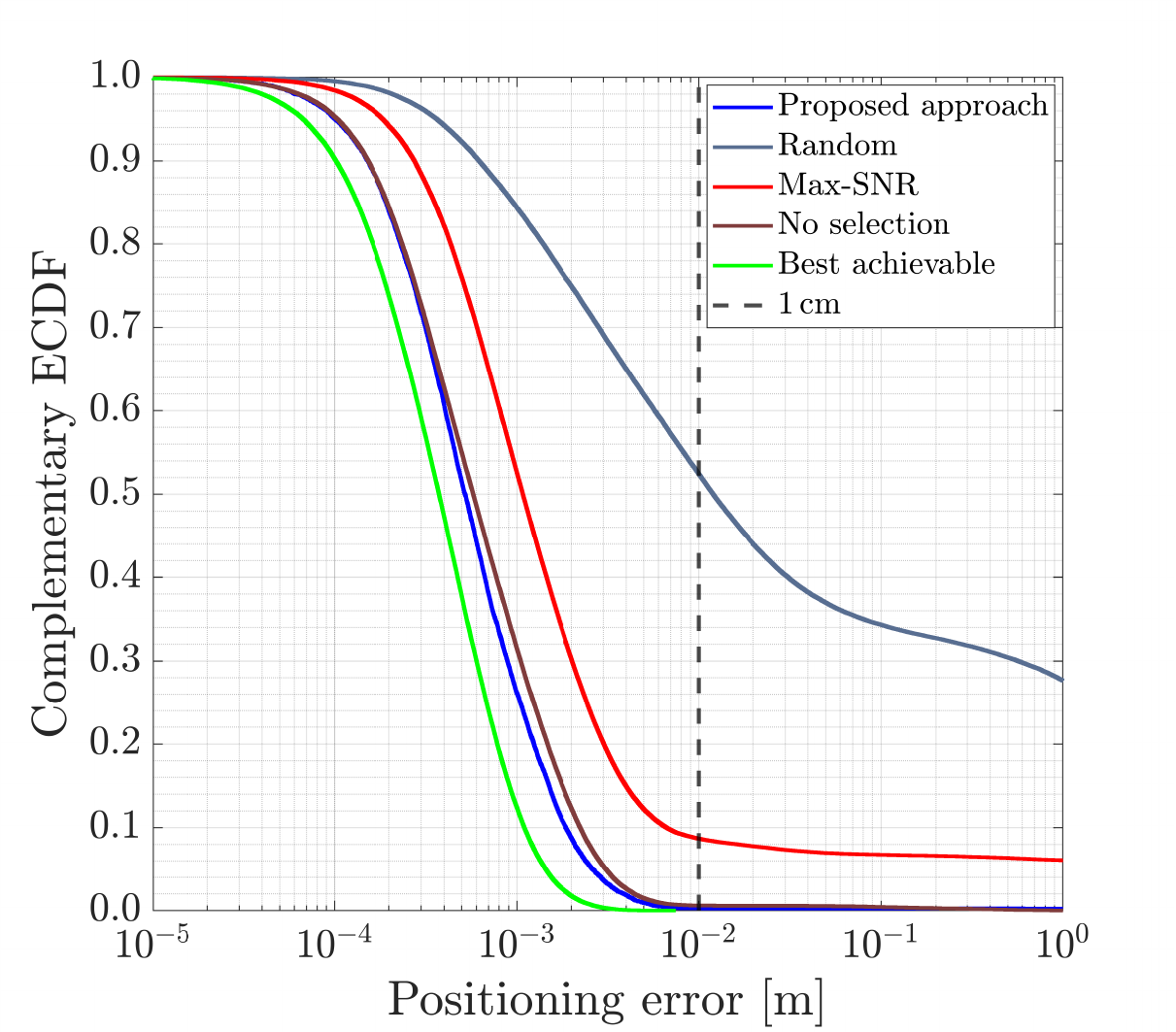}
    }
    \vspace{-2mm}
    \caption{(a) Accuracy and positioning performance with different training and test set combinations; (b)  ECDF of the proposed approach and benchmarks; (c) complementary  ECDF of the proposed approach and benchmarks.}
    \label{fig:top_three}
    \vspace{-1mm} 
\end{figure*}

\vspace{-2mm}\subsection{Differential Ambiguity Estimation Results}\vspace{-1mm}
\textls[-4]{We analyze the accuracy of the differential ambiguity estimator for all ambiguities, i.e., $\vert \mathcal{J} \vert =I-1$. The \textit{overall accuracy} is used as a performance metric. It is defined over $S$ test samples as $A_o = \frac{1}{S}\sum_{s=1}^{S} \mathbb{I}(\Delta\hat{\boldsymbol{z}}^s = \Delta{\boldsymbol{z}}^s) \times 100\%$, where $\mathbb{I}(\cdot)$ denotes the indicator function, which equals 1 when its argument is true and 0 otherwise. Here, $\Delta\hat{\boldsymbol{z}}^s$ and $\Delta{\boldsymbol{z}}^s$ represent the estimated and ground-truth differential ambiguity vectors for the $s$-th test sample, respectively. Additionally, the 95th percentile positioning error is evaluated using the \gls{gd}-based solver with all estimated ambiguities. To assess the model’s ability to adapt to different levels of phase perturbation, separate models are trained using phase perturbation standard deviation $\sigma$ selected from $\{0,\pi/200,\,\pi/100,\,\pi/50\}$ and subsequently tested on distinct datasets. The results are illustrated in Fig.~\ref{fig:acc_and_95th_error}, based on a test set containing $S=150 \times 10^3$ samples.

\textls[-12]{When both the training and test sets include phase-perturbed data, increasing the perturbation has less impact on the overall accuracy and positioning performance compared with the case where the model is trained without phase perturbations. In the phase-perturbed training scenario, the overall accuracy remains above $99\%$, and the $95$th-percentile error stays below $0.5\,\mathrm{cm}$. This demonstrates that the model can adapt to phase perturbations and still achieve highly accurate positioning.}}

\vspace{-2mm}\subsection{AP Selection Results}\vspace{-1mm}
\textls[-4]{Since \gls{ap} selection for phase-only positioning in \gls{dmimo} has not been previously evaluated, there are no directly comparable benchmark methods available in the literature. We thus compare our \gls{ap} selection method against the following four reference strategies: \emph{Random} \cite{max_snr_and_random,random_ref1}, where the pair of ambiguities is chosen uniformly at random; \emph{Max-SNR} \cite{max_snr_and_random, max_snr_ref1}, where the pair of APs having the largest \gls{snr} values is chosen for differential ambiguity estimation; \emph{No selection}, which uses all available ambiguities without AP selection; and \emph{Best achievable}, representing the ground-truth subset with two ambiguities that yields the minimum positioning error for each test sample. The perturbation parameter $\sigma$ in both the training and test sets is set to $\pi/100$. Fig.~\ref{fig:positioning_ecdf} shows the \gls{ecdf}, i.e., $P\{\mathrm{Positioning\ error}\le x\}$,
while Fig.~\ref{fig:positioning_ccdf} shows the complementary \gls{ecdf}, i.e., $P\{\mathrm{Positioning\ error}>x\}$. The $95$th and $99$th percentile errors are compared in Table~\ref{tab:perf_metrics}. The proposed \gls{ap} selection model outperforms the \textit{Max-SNR}, \textit{Random}, and \textit{No selection} benchmarks, and requires approximately $19.7\%$ fewer \glspl{flop}. Furthermore, considering the severe accuracy degradation observed in the \textit{Max-SNR} and \textit{Random} benchmarks, it can be concluded that, even when the model does not select the optimal pair, it effectively avoids the worst choices. Extremely low positioning errors such as $10~\mu m$ are observed when the AP and UE positions are perfectly measured during training; the impact of AP position uncertainties will be investigated in future work.}

\begin{table}[t!]
\centering
\caption{Positioning performance comparison of the considered AP selection models.}
\label{tab:perf_metrics}
\small 
\begin{tabular}{lcc}
\toprule
 & \multicolumn{2}{c}{\textbf{Error [cm]}} \\
\cmidrule(lr){2-3}
\multicolumn{1}{c}{\textbf{Method}} & \textbf{95th Perc.} & \textbf{99th Perc.} \\
\midrule
Best achievable   & $0.14$   & $0.23$ \\
Proposed          & $0.26$   & $0.48$ \\
No selection      & $0.30$   & $0.59$ \\
Max-SNR     & $509.8$  & $1057.4$ \\
Random  & $551.1$  & $768.7$ \\
\bottomrule
\end{tabular}\vspace{-5mm}
\end{table}

\vspace{-3mm}\section{Conclusion}\vspace{-3mm}
\textls[-2]{This paper demonstrated that carrier phase–only positioning, when combined with \gls{dl}–based solutions, can achieve high-precision localization in realistic \gls{dmimo} deployments. By explicitly addressing phase synchronization errors, the proposed \gls{dlmf} framework enhances robustness against practical impairments while retaining centimeter-level accuracy. Moreover, the integration of adaptive AP selection strategies ensures efficient use of network resources. These results highlight the potential of phase-only positioning and learning-driven approaches as key enablers for reliable and accurate localization in future 6G systems. Future work will extend the proposed
framework to cover multipath and three-dimensional scenarios.}

\section{ACKNOWLEDGMENTS}
This work was supported by the MiFuture project under the HORIZON-MSCA-2022-DN-01 call (Grant number: 101119643), by the SNS JU project 6G-DISAC under the EU's Horizon Europe research and innovation Program under Grant Agreement No 101139130, and by the Swedish Research Council (VR) through the project 6G-PERCEF under Grant 2024-04390. The work was also in part supported by Business Finland under the 6G-ISAC project, and in part by the Research Council of Finland under the grant 359095.


\bibliographystyle{IEEEtran}
\bibliography{refs}

\end{document}